# Effectiveness of Multi-Physics Numerical Model in Simulating Accelerated Corrosion with Spatial and Temporal Non Uniformity

Shanmukha Shetty[1], Sauvik Banerjee[2*], Siddharth Tallur[3], Yogesh M. Desai[4]

[1,2,4]*Department of Civil Engineering, Indian Institute of Technology Bombay, Powai, Mumbai, 400076, India*

[3]*Department of Electrical Engineering, Indian Institute of Technology Bombay, Powai, Mumbai, 400076, India*

*Corresponding Author, E-mail address:* sauvik@civil.iitb.ac.in

**Abstract**

This study is motivated by the need to develop an efficient numerical model to simulate non-uniform interfacial degradation of reinforcing steel in concrete in an accelerated corrosion setup. In this study, a multi-physics finite element (FE) model is presented that takes into consideration the spatial and temporal non-uniformity of corrosion induced degradation in rebar, and eliminates the assumption of uniform mass loss and its linear variation with time as per available literature that uses classical approach of Faraday's law. The model is validated experimentally with accelerated corrosion setup designed to induce partial corrosion. Further, the possibility of extending this model to monitor natural corrosion is discussed with required modifications. Unlike previous studies, pore saturation (PS) is continuously monitored and its existing experimental correlations with electrolyte conductivity and oxygen diffusivity in the vicinity of partial corrosion are adopted so that the model can be extended to simulate natural corrosion. These evaluations can be made completely nondestructive and in real time to capture the influence of local environment. The proposed methodology also captures the effect of differential aeration pertaining to local exposure. Therefore, the challenges in incorporating influence of local environment by the use of alternative parameters such as relative humidity from real climate change predictions are eliminated. It is shown that the multi-physics model is effective and convenient to simulate the non-uniform time dependent mass loss with acceptable accuracy and that its capability can be extended to monitor natural non uniform corrosion on a space-time frame.

## 1 Introduction

The effectiveness of service life predictions in reinforced concrete (RC) structures subjected to corrosion induced deterioration can be enhanced with accurate corrosion monitoring and strength degradation evaluation strategies. As a step towards this, numerical studies [1-10], and analytical corrosion models [11-13] coupled with service life prediction models [14-19] were developed. Experimental investigations were also performed to examine the efficacy of the numerical and analytical studies [20,21]. Researchers [22] have developed a stochastic computational framework to incorporate various sources that result in uncertainties in predicting the performance of structures. However, integration of these models with real-time investigations specific to local exposure by considering the non-uniform nature of influencing parameters was not attempted.

Only few investigations had considered real-time corrosion monitoring incorporating nondestructive evaluations [23-29], but are limited to monitoring corrosion initiation and propagation assuming uniform mass loss and linear rate of corrosion with no possibility of obtaining the non-uniform degraded geometry. These idealizations stem from the challenges involved in incorporating coupled interactions between structure and its local environment on a space-time frame in a 3D domain.

Researchers [30] envisioned that the next generation strategies to maintain RC structures will combine multi-physics numerical simulation having multi-scale capabilities with in-situ monitoring, allowing continuous updating of what is called as 4D simulation of structural performance. This forms the basis of present research work.

Recently, a 2D numerical model simulating non uniform corrosion incorporating the influence of temperature and relative humidity (T-RH) from real climate with temporal variations was presented in [31,32]. These studies concluded that, sampling methods of real climate data to simplify the real variability, affect the corrosion initiation and propagation rate showing higher corroded depth in case of daily variations in comparison with weekly and monthly variation schemes. It is evident from these observations that the uncertainties associated with real climate data can seriously affect corrosion rate predictions because of fluctuations in T-RH, which was also identified in [33]. Therefore, a probabilistic approach to account for these uncertainties was adopted, but the real-time local exposure and the environment of member under investigation were ignored.

Thus, the present study proposes a real time data based model to arrive at conductivity characteristics, oxygen diffusivity and moisture (pore saturation) which directly influences the corrosion current rather than using T-RH from climate data [34-38].

While modelling capabilities of COMSOL Multi-physics® were explored by few researchers [39,40], in the present study a fully coupled model is developed and analyzed in COMSOL Multi-physics®. Local corrosion current densities are evaluated in order to obtain degradation profiles of reinforcing bars after suitably modifying the mathematical model for electrode deposition or dissolution proposed in [41–43]. Transport of diluted species, which simulates the effect of differential aeration with oxygen diffusion is also considered. The presented multi-physics model is validated for accelerated corrosion setup [44]. The model can be extended to simulate naturally corroding structures by adopting the PS and electrolyte conductivity data obtained by scheduled monitoring using concrete moisture meter and four probe resistivity meter respectively and oxygen diffusivity as a function of PS can be obtained using experimental relationship presented in [35,36]., Also the ingress of chloride can be modelled as transport of diluted species in multi-physics modelling environment. Extending the application to model natural corrosion will form the scope of future investigations and is briefly mentioned in the present study as one of the applications and adaptability of the proposed model.

## 2 Numerical modelling of corrosion propagation phase in accelerated corrosion setup

A multi-physics FE model coupling transport of chemical species with electrochemical reactions of corrosion to simulate accelerated partial corrosion with impressed current technique in RC members is developed. The model also takes into consideration the local environment of experimental setup using COMSOL Multi-physics®. However, for real time monitoring of naturally corroding structures the chloride ingress needs to be incorporated as a transport of diluted species physics in modelling. Alternatively, many numerical models are also available from literature [5,31,32,45–49] to arrive at the non-uniform chloride profiles. The threshold chloride concentration is made readily available in accelerated corrosion experimentation and therefore, the model represents the propagation phase of corrosion wherein the incorporation of chloride ingress is not relevant.

2.1 Theoretical background of multi-physics modelling for corrosion

The multi-physics model uses FE Method to solve the partial differential equations representing the oxygen diffusion with corrosion propagation phase. The chemical species transport physics (nomenclature used in COMSOL Multi-physics®) is used to model the diffusion of oxygen, and the coefficient form PDE in mathematics section of COMSOL Multi-physics® is adopted to simulate the Nernst plank secondary current distribution governing the conductivity and potential of electrode, and electrolyte surrounding the rebar which is the concrete itself. The model automatically predicts the anodic and cathodic locations that are developed due to differential aeration [50], and their distributions over time.

*2.2.1 Corrosion reaction kinetics*

In non-uniform wet corrosion, which is the most common type of chloride induced corrosion, an electrochemical cell is setup forming anode and cathode on the corroding section. Many reactions are possible during corrosion, and are mainly categorized as oxidation at anode and oxygen reduction at cathode. The reactions in an impressed current setup are considered from [4].
In corrosion propagation phase, especially in case of chloride induced corrosion macro-cells are formed [31]. The macro-cells may also form because of heterogeneous steel-concrete interface and may also arise from the differences in electrode potentials resulting from differential aeration. Micro-cell results in uniform corrosion without any net flow of current through electrolyte [50].

$$i_{mac,a} = i_a - i_c \tag{1}$$

$$i_{mac,c} = -i_c \tag{2}$$

$$i_{mic,a} = i_c \tag{3}$$

$$i_{corr} = i_{mac,a} + i_{mic,a} \tag{4}$$

where $i_{mac,a}$ and $i_{mac,c}$ are macro-cell current densities at anode and cathode respectively (A/m$^2$), $i_{mic,a}$ is micro-cell current density at anode (A/m$^2$), $i_{corr}$ is the total corrosion current density (A/m$^2$), which in the absence of micro-cell corrosion, becomes $i_{mac,a}$.

*2.2.2 Modelling oxygen transport*

The diffusion process is simulated using Fick's first law. The corrosion current at cathode depends on rate of corrosion at cathode, which is governed by availability of oxygen at cathodic surface. On the other hand, the concentration of oxygen at cathodic surface depends on the molar flux evaluated from the corrosion current. This give rise to a cyclic coupling and is addressed in the modelling.

$$\frac{\partial C_{O2}}{\partial t} = \nabla \cdot (D_{O2} \nabla C_{O2}) \tag{5}$$

where $D_{O2}$ is effective diffusion coefficient of oxygen in concrete (m²/s), $C_{O2}$ is the concentration of oxygen in concrete (mol/m³).
The boundary condition at cathodic surface is obtained by using Faraday's law of electrolysis on equating oxygen flux to cathodic current density as shown in Eq. (6).

$$\frac{i_{O2}}{Fz} = -D_{O2}(\boldsymbol{n}.\nabla C_{O2}) \tag{6}$$

where $i_{O2}$ is the cathodic current density; F (96485.3365 C/mol) is Faraday's constant; z is electrons involved in reaction.

An initial constant concentration of $C_{O2}^{ref}$ which is the average atmospheric concentration of oxygen equal to 8.6 mol/m³ is considered on all exposed concrete surfaces.
Zero flux boundary condition is adopted on all other surfaces as shown in Eq. (7).

$$\boldsymbol{n}.(D_{O2}\nabla C_{O2}) = 0 \tag{7}$$

*2.2.3 Modelling charge transport*

In the presence of an electric field, the ions dissolved in pore solution serve as charge carriers, the difference in electric potential ϕ (corrosion potential) in concrete generates the electric field. As the process of corrosion is electrochemical in nature, charge transport in concrete needs to be considered. The governing equation for charge transport at steady state is given by Eq. (8).

$$\nabla.\frac{1}{\rho}(\nabla\phi) = 0 \tag{8}$$

Here $\rho$ is electrical resistivity of concrete (Ω.m), the reciprocal of which gives electrical conductivity σ (s/m). $\rho$ is used to simplify the phenomenon of complex ionic transport and can be obtained experimentally or by relating with PS [34]. At steel-concrete interface both anodic and cathodic reactions are specified with corresponding current densities on the entire rebar surface., This macro-cell current $i_{mac}$ is equated to electrolyte potential flux, thus defining boundary conditions at anodic and cathodic sites on full length of rebar embedded in concrete. Even though, the entire surface of rebar is specified with anodic and cathodic reactions, the differential aeration specified automatically facilitates the anodic and cathodic sites to evolve naturally.

$$i_{mac} = \sum_m \boldsymbol{i_{loc,m}} = i_a - i_c \tag{9}$$

$$i_{mac} = \boldsymbol{n}.\boldsymbol{i_l} = \boldsymbol{n}.(-\sigma\nabla\phi) \tag{10}$$

Here $i_a$ and $i_c$ are anodic and cathodic current densities respectively (A/m²), $\boldsymbol{i_l}$ is the current density of electrolyte (A/m²), m is tag of electrode reactions defined, $\boldsymbol{i_{loc,m}}$ is the local current density for a particular reaction m (A/m²). Zero flux boundary conditions are enforced on all surfaces other than anodic and cathodic locations.

$$\boldsymbol{n}.(-\frac{1}{\rho}\nabla\phi) = 0 \tag{11}$$

$i_a$ and $i_c$ the anodic and cathodic current densities (A/m²) are evaluated using Eq. (12)) and (Eq.(13) respectively [38]. The polarization experimentation results were adopted from [34].

$$i_a = i_a^0 \times 10^{\frac{\eta_a}{A_{Fe}}} \tag{12}$$

$$i_c = \frac{C_{O2}}{C_{O2}^{ref}} \times i_c^0 \times 10^{\frac{\eta_c}{A_{O2}}} \tag{13}$$

Here $i_a^0$ and $i_c^0$ are exchange current densities for anodic and cathodic reactions respectively (A/m²), $C_{O2}$ is the concentration of oxygen diffused in concrete domain (mol/m³), $C_{O2}^{ref}$ reference or atmospheric oxygen concentration (8.6 mol/m³), $A_{Fe}$ and $A_{O2}$ are anodic and cathodic tafel slopes respectively, $\eta_a$ and $\eta_c$ are anodic and cathodic over-potentials, respectively and are given by Eq.(14) and Eq.(15).

$$\eta_a = E_{appl,a} - \phi - E_{Fe}^{eql} \tag{14}$$

$$\eta_c = E_{appl,c} - \phi - E_{O2}^{eql} \tag{15}$$

Here ϕ is the electric potential at the interface (Volts), $E_{appl,a}$ and $E_{appl,c}$ are the external voltage applied in accelerated corrosion experiment at anode (Rebar) and cathode (galvanized wire mesh wrapped on concrete surface on corroding section immersed in NaCl) respectively (Volts), $E_{Fe}^{eql}$ and $E_{O2}^{eql}$ are the equilibrium potential for anode and cathode reactions respectively (Volts). The constants and tafel slopes are extracted from literature [34]. Initially, corrosion potential ϕ is evaluated in the domain as function of time by solving Eq. (8), which takes the form of Laplace's Equation. Using the corrosion potential ϕ, anodic and cathodic current densities are evaluated, that can further be used in evaluation of micro and micro-cell current densities. From these parameters total current densities at any point on the surface of corroding rebar can be evaluated.

*2.2.4 Application of Faraday's law in evaluating metal dissolution*

Faraday's laws of electrolysis in combined form and its variations are very often used in evaluating weight loss w, and dissolution rate commonly referred to as rate of corrosion.

$$w = \frac{q_m M}{nF} \tag{16}$$

Here $q_m$ is the electric charge (C), responsible for corrosion associated with electrode reaction m, M is the molar mass (kg/mol), n is the number of electrons transferred and F is the Faraday's constant (F = 96485.3365 C/mol).

Electrical charge, q, is the integral of current **i** over time for which the charge is passed,

$$q = \int_{t1}^{t2} \boldsymbol{i}\, dt \tag{17}$$

and the current density **j** is defined as the current per unit electrode area (A/m²), i.e. the surface area of that portion of electrode participating in electrolysis due to corrosion [43]

$$\boldsymbol{j} = \frac{i}{A} \tag{18}$$

Therefore, $q = \int_{t1}^{t2} A\boldsymbol{j}\, dt \tag{19}$

The use of this expression (Eq. (19)) is preferred than (Eq. (17)) as the area of corroding surface and **i** continue to change with corrosion progress due to dissolution or deposition.

## 2.3 Non-uniform rate of dissolution using numerical model

The multi-physics model facilitates evaluation of point wise time dependent local current densities responsible for corrosion. Therefore, there is a provision for adding arbitrary number of dissolving-depositing species. The surface concentration variables of dissolving-depositing species can be used to calculate thickness of dissolution or deposition and the rate of which is used to set the boundary velocity. The velocity of dissolution is expressed as [41]

$$\frac{\partial \mathbf{x}}{\partial t} \cdot \mathbf{n} = \mathrm{v}_{dis,\,tot} \qquad (20)$$

Here $\mathrm{v}_{dis,\,tot}$ is the total dissolution velocity (m/s), defined as the sum of velocity contributions for all species and electrode reactions, and is expressed as:

$$\mathrm{v}_{dis,\,tot} = \sum_i \frac{M_i}{\rho_i} \sum_m \frac{v_{i,m} i_{loc,m}}{n_m F} \qquad (21)$$

where $i_{loc,m}$ is the local current density or corrosion current density **j** at that point where dissolution is intended to be evaluated (A/m$^2$). $v_{i,m}$ is the stoichiometric coefficient of species i with respect to reaction m, $\rho_i$ is the density of metal undergoing corrosion, $n_m$ is the number of electrons transferred.

## 3 Validation of numerical model

The experimental setup of accelerated corrosion with impressed current proposed by [44] for analyzing pitting corrosion of rebar embedded in concrete is considered for inducing partial corrosion. The details of the experimental setup can be found in [44]. In this work, impressed current is applied through a DC power supply with constant voltage of 25 V maintained over a period of 12 days, and the corresponding daily variation of current is measured. The middle RC portion of 150mm length is immersed in NaCl solution to introduce partial corrosion. The numerical model simulates this experimental setup of accelerated corrosion with impressed current, and the mass loss comparison is considered for validation.

## 4 Specifications of Multi-physics finite element modelling to simulate accelerated corrosion

Mesh descritisation depends on the type of physics being solved and must ensure convergence. Since the numerical model adresses two physics namely, current distribution using coefficient form PDE under mathematics node and transport of diluted species, each having a different type of meshing requirements, the physics conrolled meshing capabilities of COMSOL Multi-physics® is adopted, that predominantly uses tetrahedral mesh as shown in Fig.4(b),Fig.4(c) and Fig.5(a). Mesh convergence study is carried out to finalize the meshing criteria.

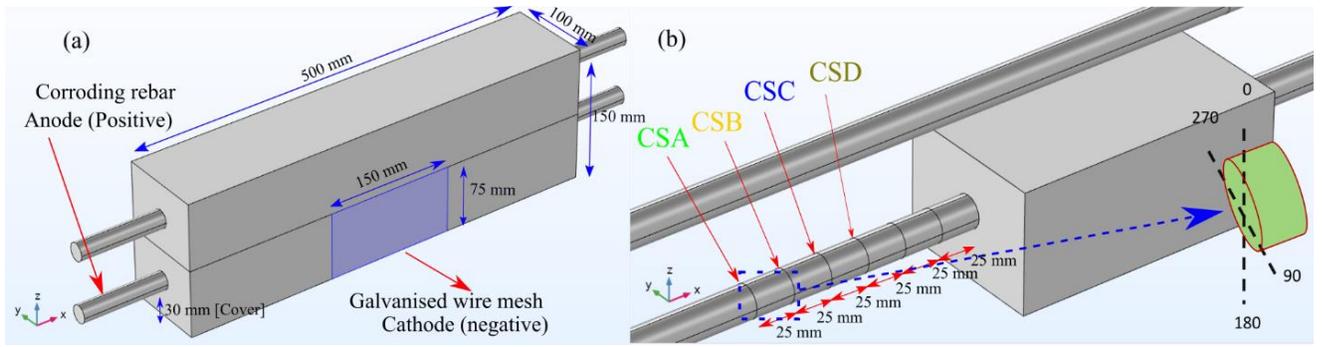

**Fig. 1** (a) Numerical model of accelerated corrosion setup and (b) Corroding length of rebar with typical cross sections considered for analysing deformed geometry

4.1 Input parameters for simulating electrochemistry of corrosion in the numerical model

**Table 1** Input Parameters for Simulating Electrochemistry of Corrosion in Numerical Model [34,40]

| Notation | Value | Units | Description |
| --- | --- | --- | --- |
| $A_{Fe}$ | 0.41 | V | Tafel slope in iron oxidation |
| $A_{O2}$ | -0.18 | V | Tafel slope in oxygen reduction |
| $E_{Fe}^{eql}$ | -0.76 | V | Iron oxidation equilibrium potential |
| $E_{O2}^{eql}$ | 0.189 | V | Oxygen reduction equilibrium potential |
| $i_a^0$ | 7.1E-5 | A/m² | Iron oxidation exchange current density |
| $i_c^0$ | 7.7E-7 | A/m² | Oxygen reduction exchange current density |
| $E_{appl,a}$ | 25 | V | Applied potential to rebar |
| $i_{a\,mesh}^0$ | 1E-7 | A/m² | Oxidation exchange current density on mesh |
| $i_{c\,mesh}^0$ | 1E-4 | A/m² | Oxygen reduction exchange current density on mesh |
| $E_{Fe,mesh}^{eql}$ | -0.44 | V | Mesh oxidation equilibrium potential |
| $E_{o2,mesh}^{eql}$ | 0.189 | V | Oxygen reduction equilibrium potential |
| $A_{O2,mesh}$ | -0.25 | V | Oxygen reduction tafel slope on mesh |
| $A_{Fe,mesh}$ | 0.12 | V | Mesh oxidation tafel slope |
| $E_{appl,c}$ | 0 | V | Applied potential to mesh |

4.2 Oxygen diffusivity and electrolyte conductivity as a function of pore saturation

The oxygen diffusivity and concrete resistivity are related to the PS levels of concrete. PS level has its direct bearing on temperature and humidity. Therefore, if PS is evaluated as a function of time, then empirical relations amongst PS, oxygen diffusivity and concrete resistivity presented in [34-37] can be used to simulate effect of local environment. However, in natural corrosion PS can be evaluated using NDT as described in [51–53]. For the water cement ratio and cover thickness considered in the present model, the relation among PS, electrolyte conductivity and oxygen diffusivity are adopted from [34], that are deduced by direct interpolation from the experimental investigations of [35-37], and therefore, the oxygen diffusivity as a function of PS for a given water cement ratio and cover thickness can be obtained by interpolation [35,36] and electrical conductivity from [37]. Three cubes of M25 grade are casted and cured for 28 days. These cubes are then immersed in NaCl solution with same concentration used in [44], and daily water absorption is monitored for 10 days to calculate PS as presented in Table 2. It is assumed that the PS in experimentation followed a similar profile. However, because of the first crack in concrete on 11th day, there was a spike in water ingress as evident from impressed current variations [44]. Therefore, a PS of 0.7 is assumed on 11th and 12th day based on the similarities in trend of variation of PS and impressed current [44].

**Table 2** Electrolyte Conductivity and Oxygen Diffusivity as a function of PS [34]

| PS | Days | Electrolyte Conductivity ($\sigma$, S/m) | Effective Diffusivity of $O_2$, ($D_{O2}$, m$^2$/s) |
| --- | --- | --- | --- |
| 0.200 | 0 | 1.75E-04 | 1.52E-08 |
| 0.449 | 1 | 3.42E-03 | 6.62E-09 |
| 0.606 | 2 | 7.16E-03 | 2.70E-09 |
| 0.700 | 3 | 9.82E-03 | 1.50E-09 |
| 0.699 | 4 | 9.77E-03 | 1.51E-09 |
| 0.688 | 5 | 9.38E-03 | 1.62E-09 |
| 0.658 | 6 | 8.29E-03 | 1.92E-09 |
| 0.626 | 7 | 7.53E-03 | 2.39E-09 |
| 0.621 | 8 | 7.44E-03 | 2.47E-09 |
| 0.607 | 9 | 7.17E-03 | 2.69E-09 |
| 0.605 | 10 | 7.14E-03 | 2.72E-09 |
| 0.700 | 11 | 9.80E-03 | 1.50E-09 |
| 0.700 | 12 | 9.80E-03 | 1.50E-09 |

**5 Methodology developed to quantify metal dissolution**

The methodology adopted in evaluating weight loss from experiment is shown in Fig.2. Note that in the experimental setup a constant voltage of 25V is applied with regular monitoring of current, as a result current varies as a function of time due to various factors of reaction kinetics and the environment of experimental setup, therefore mass loss using Faradays law may be calculated on a daily basis and cumulative weight loss can be evaluated, but the weight loss cannot be predicted beforehand to plan on the experiment. However, in the proposed numerical model local current density evaluated point wise is then used to obtain non-uniform point wise corrosion rate (dissolution velocity), from which, pointwise dissolution is evaluated by multiplying dissolution velocity with a small time interval dt for which the dissolution velocity is assumed to be constant, which finally can be used in modelling deformed rebar geometry to calculate weight loss as depicted in Fig.3.

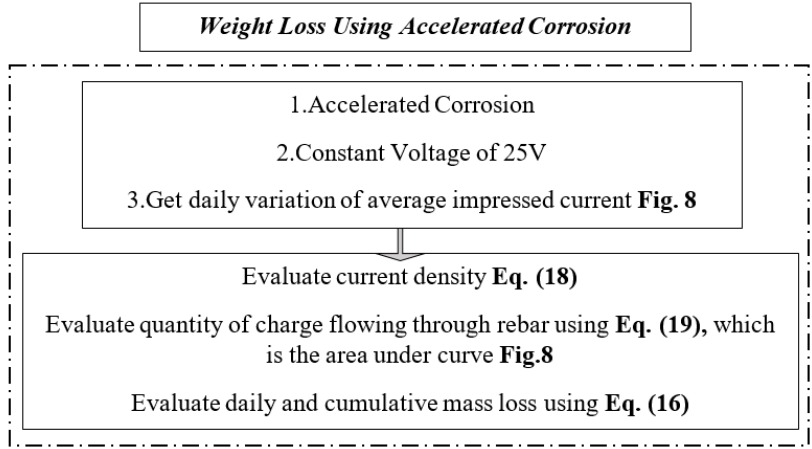

**Fig. 2** Flowchart for evaluating mass loss using accelerated corrosion experiment

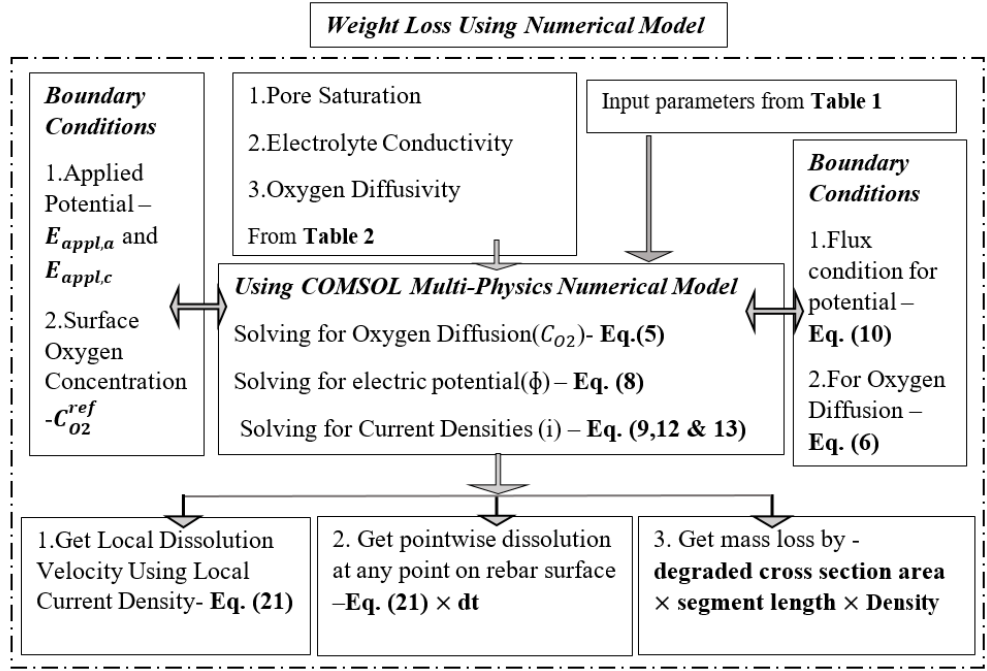

**Fig. 3** Flowchart for modelling corrosion induced degradation with space time non uniformity using multi-physics numerical model

## 6 Results and discussion

The results are evaluated at half section of length 75mm of the total corroding portion of length 150mm using symmetry in the model shown in Fig. 1. This portion of 75mm length is divided into four typical cross sections namely CSA, CSB, CSC, and CSD. CSA and CSD segments are of length 12.5mm, whereas CSB and CSD segments are of length 25mm. The results are evaluated at $10^0$ angualar spacing on the circumference of each cross section and is assumed to be constant over the length of that particular segment for evaluating deformation profile and weight loss as shown in Fig. 4. However, the model can evaluate all these parameters at any point. The evaluated electrolyte potential and the corresponding total electrolyte current density shown in Fig.5 (b) represents the the locations where corrosion is severe, and can be seen in further results of degraded geometry. These observations are in concorrence with the experimental observations [44].

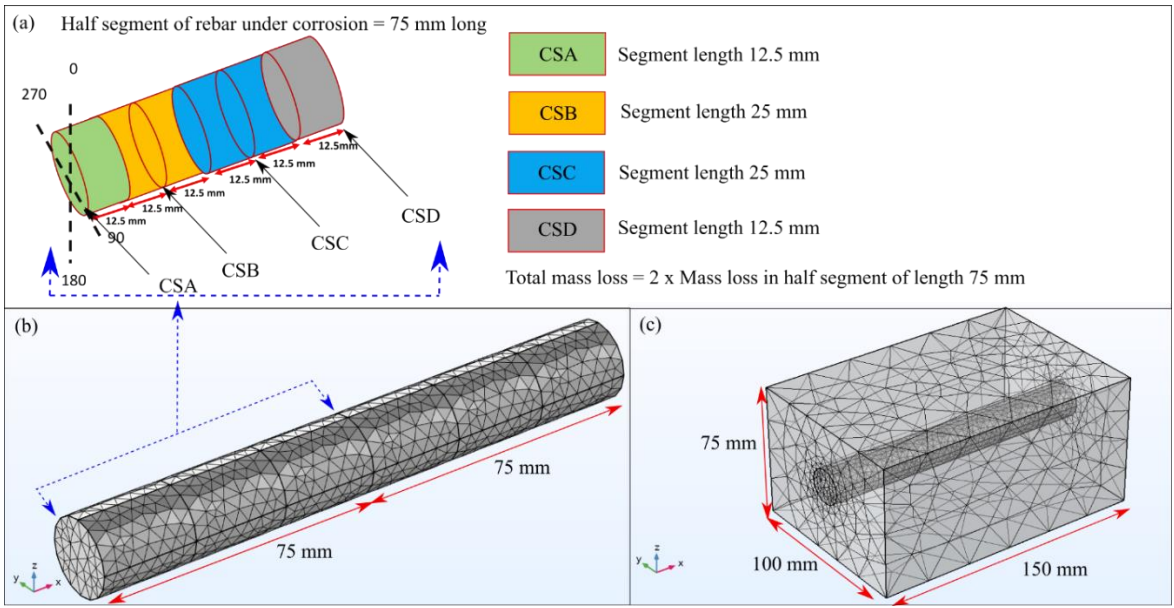

**Fig. 4** (a) Length of each segment, (b)Corroding rebar with meshing, and (c) Corroding unit immersed in NaCl solution

It may be noted that as the central portion of rebar is deprived of oxygen, this location develops into anodic site and adjacent location on the same rebar which has relatively larger supply of oxygen acts as cathodic site. This development of anodic and cathodic macro cells on the same rebar surface is completely ignored in estimating mass loss using Faraday's law from accelerated corrosion experiment, based on the assumption that the entire rebar subjected to impressed current will act as anode. The electrode and electrolyte current density in Fig.5 (c), follow similar trend that are evaluated from corresponding potentials.

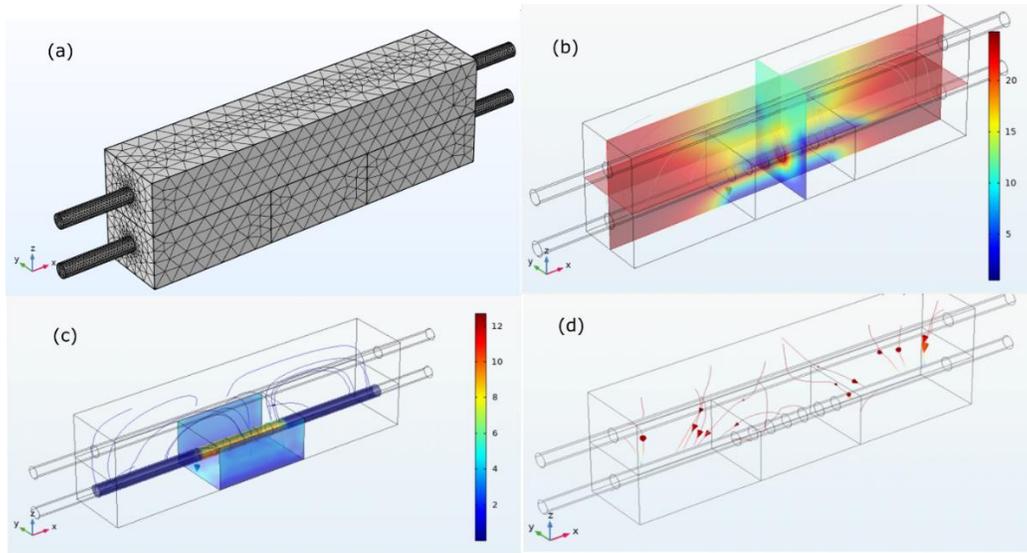

**Fig. 5** (a) Specimen for simulation (b), Electrolyte and Electrode potential distribution for 12$^{th}$ day, (c) Electrolyte current density vector and Total electrode current density at surface for 12$^{th}$ day and (d)Total flux due to consumption of oxygen diffused throgh concrete at 12$^{th}$ Day

Fig. 5(d) represents the diffusion of oxygen governing the formation of anodic and cathodic sites, which is not taken for consideration in evaluating weight loss location using conventional approach of Faradays law in an impressed current setup. This could be one of the reasons for the differences between predicted weight loss and experimental results as observed in previous studies [54,55], which suggested the need for further investigation to identify the cause for deviation. But the consideration of diffusion of oxygen would help in reducing the deviation that can be further seen in weight loss comparisons. In modeling, the oxygen diffusivity is allowed through exposed portions of concrete surface with an equilibrium concentration of 8.6 mol/m$^3$, which is the concentration of oxygen in air. Also the portion that is immersed in NaCl is deprived of oxygen exposure. It can be clearly seen from Fig. 5(d) that the adjacent portions to the left and right side of middle corroding portion of rebar is consuming oxygen, therefore, embeded rebar here is potentially acting as cathode.

6.1 Local Current Density

At every $10^0$ intervals on CSA, CSB, CSC, and CSD sections the point wise corrosion current density is obtained from the numerical model as an average daily data for 12 days of corrosion. The daily data is taken as an average over 24 hours. The simulation was carried for 12 days as the crack started appearing on the 11$^{th}$ day in accelerated corrosion experiment as observed in [44]. The trend of local current density shown in Fig.6 is along similar lines with the electrolyte potential and total current densities, and shows that maximum local current density is at the bottom portion of the rebar. The possible reason for this could be the closest proximity of this portion with the galvanised wire mesh resuting in quicker transaction of ions. Another reson that can be found from the experimental setup is that this corroding portion of rebar recieves NaCl at a much faster rate than other sections due to lesser cover thickness needed to be travelled for NaCl. Further, it can be observed that at CSA as in Fig.6(a), the maximum current density is at $180^0$. But in CSB, CSC and CSD, in Fig.6(b), Fig. 6(c), and Fig.6(d) respectively, the maximum occurs not exactly at $180^0$ but adjacent to it. This is possibly because of their locations, which are within the corroding unit immersed in NaCl ,whereas, CSA is at the junction of corroding and non corroding unit and is subject to diffusion of oxygen from left unit of concrete exposed to surface. Also, at CSB,CSC and CSD, the sections from $0^0$ to $90^0$ and $270^0$ to $0^0$ are belonging to the unit of concrete exposed to oxygen and not immersed in NaCl. But, the remaining portion of the rebar is not exposed to oxygen and immersed in NaCl. Thus, for CSB, CSC and CSD, the $90^0$ and $270^o$ locations have oxygen availability faster than $180^0$ location, but $180^0$ location has faster supply of NaCl solution, but no or very less oxygen supply.

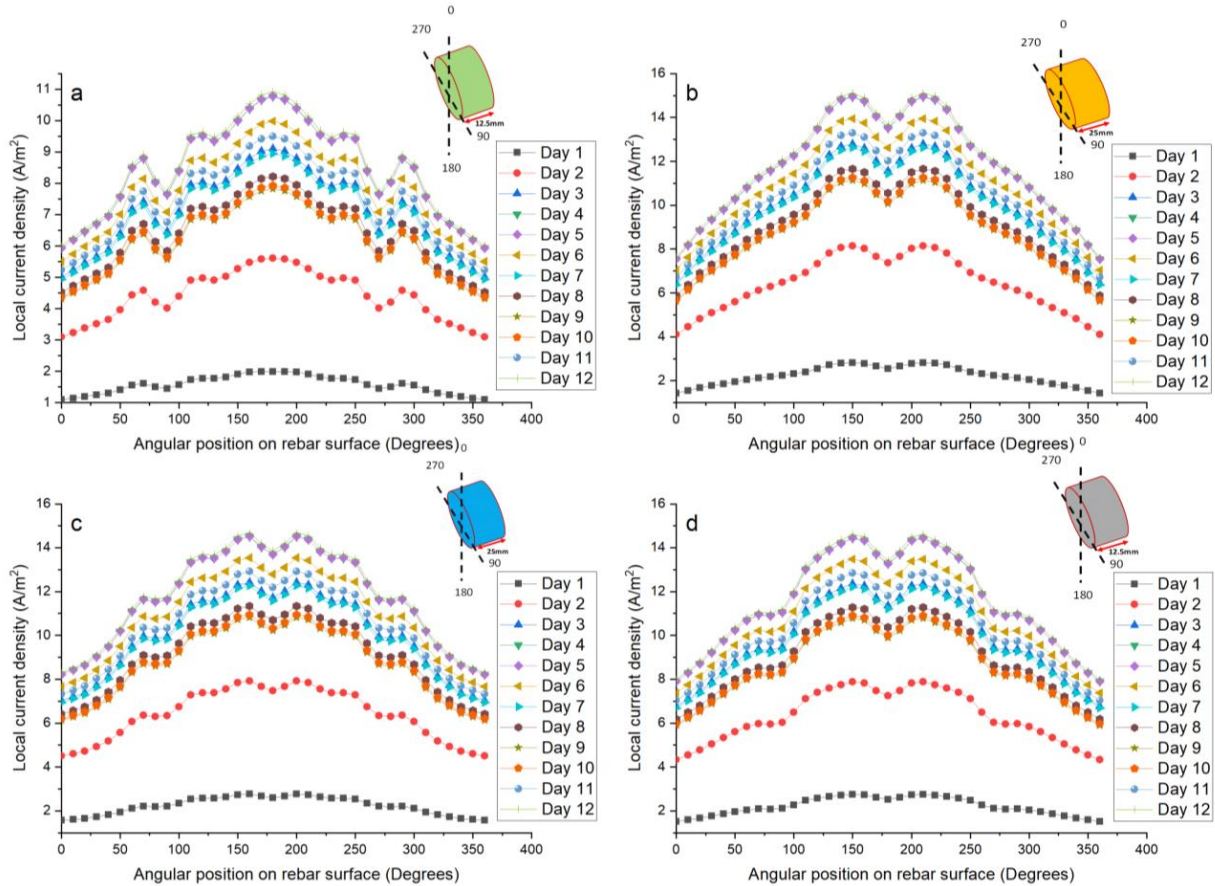

**Fig. 6** Daily average local current density with time at (a) CSA, (b) CSB, (c) CSC and (d) CSD

Thus, the locations that are between $90^0$ and $180^0$ as well as between $180^0$ and $270^0$ has a better availability of combination of NaCl solution and Oxygen diffusion.

6.2 Profile of corroded rebar geometry

Point wise degraded geometry is evaluated using local current densities. The profile of degraded geometry follows the profile of local current density, and the lower section of rebar gets corroded the most following the same reasoning as provided in section 6.1. This approach can effectively simulate non-uniformity of corrosion and resulting degradation profile as presented in Fig. 7.

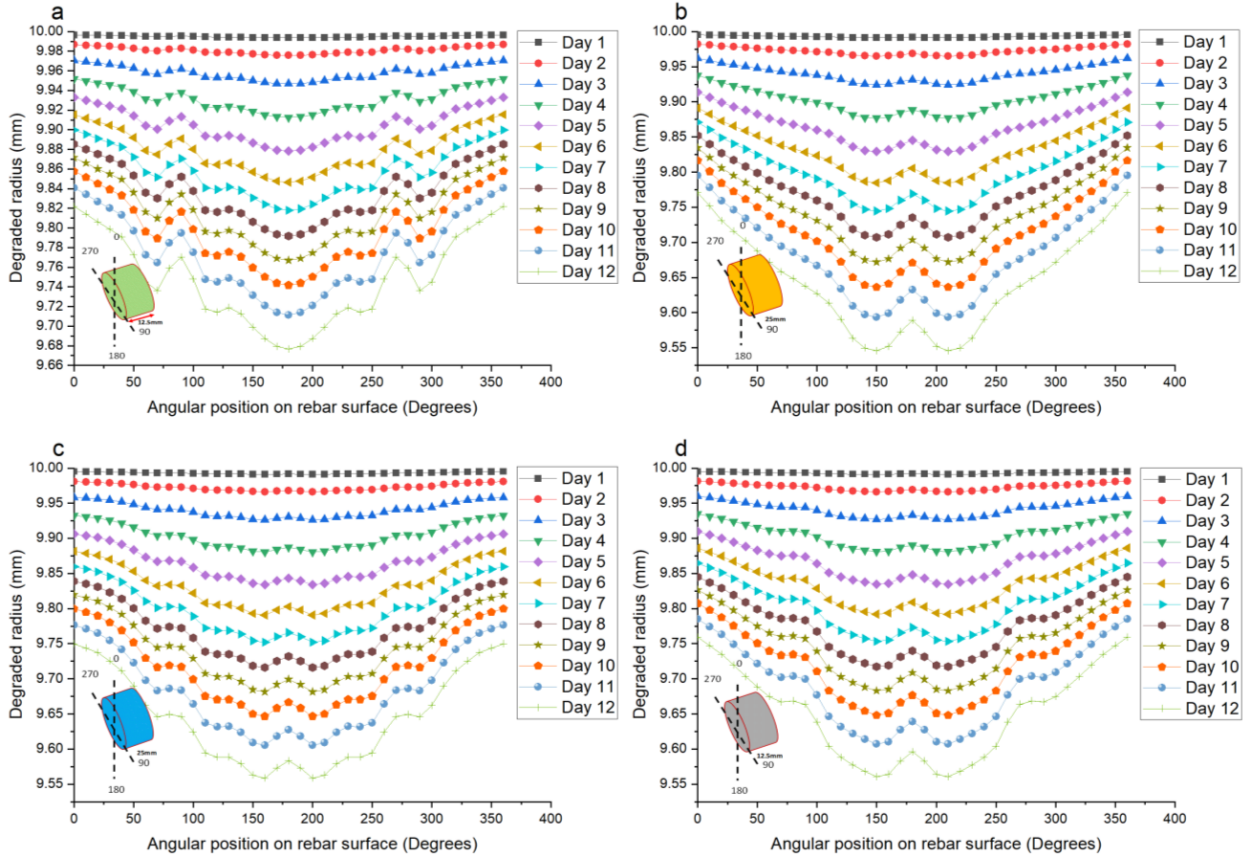

**Fig. 7** Degraded rebar profile due to corrosion at (a) CSA, (b) CSB, (c) CSC and (d) CSD

6.3 Evaluation of mass loss from experimental investigation

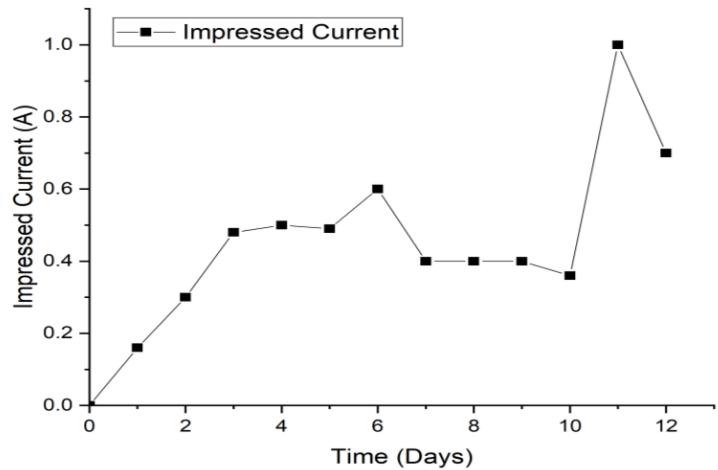

**Fig. 8** Daywise Impressed Current measured in accelerated corrosion experiment [44]

A much better way of using Faraday's law in evaluating mass loss from experiment is to use varying impressed current I, and use trapezoidal rule to evaluate the total charge flow (area under the curve in Fig.8) in Faraday's law of electrolysis for each time interval that is considered (24 hours) from Fig. 8 [44]. Table 3 shows the mass loss evaluated using average current $I_a$, considering I to be constant for 24 hours.

**Table 3** Mass Loss Evaluated from Experimental Investigation

| V (Volts) | I (A) | $I_a$ (A) | Mass Loss Per Day (g) | Cumulative Mass Loss (g) |
|---|---|---|---|---|
| 0 | 0 | 0 | 0.00 | 0.00 |
| 25 | 0.16 | 0.16 | 4.00 | 4.00 |
| 25 | 0.3 | 0.23 | 5.75 | 9.75 |
| 25 | 0.48 | 0.39 | 9.75 | 19.50 |
| 25 | 0.5 | 0.49 | 12.25 | 31.75 |
| 25 | 0.49 | 0.495 | 12.38 | 44.13 |
| 25 | 0.6 | 0.545 | 13.63 | 57.75 |
| 25 | 0.4 | 0.5 | 12.50 | 70.25 |
| 25 | 0.4 | 0.4 | 10.00 | 80.25 |
| 25 | 0.4 | 0.4 | 10.00 | 90.25 |
| 25 | 0.36 | 0.38 | 9.50 | 99.75 |
| 25 | 1 | 0.68 | 17.00 | 116.75 |
| 25 | 0.7 | 0.85 | 21.25 | 138.00 |

6.4 Evaluation of mass loss from numerical model

The 3D deformed geometry is modelled in CAD, to obtain volume of corroded rebar and multiplying it with density of steel gives weight loss and is depicted in Table 4.

**Table 4** Mass Loss Evaluated from Numerical Model

| | Area loss from CAD model (mm²) | | | | Mass Loss Per Day (g) | Cumulative Mass Loss (g) |
|---|---|---|---|---|---|---|
| Day | CSA | CSB | CSC | CSD | | |
| 0 | 314.159 | 314.159 | 314.159 | 314.159 | 0.00 | 0.00 |
| 1 | 312.252 | 312.120 | 312.120 | 312.125 | 2.37 | 2.37 |
| 2 | 311.370 | 310.834 | 310.840 | 310.861 | 3.80 | 6.17 |
| 3 | 309.949 | 308.820 | 308.842 | 308.888 | 6.03 | 12.20 |
| 4 | 308.274 | 306.475 | 306.516 | 306.592 | 8.63 | 20.83 |
| 5 | 306.602 | 304.139 | 304.199 | 304.304 | 11.23 | 32.06 |
| 6 | 305.057 | 301.966 | 302.044 | 302.176 | 13.64 | 45.70 |
| 7 | 303.674 | 300.008 | 300.100 | 300.256 | 15.81 | 61.51 |
| 8 | 302.407 | 298.204 | 298.309 | 298.487 | 17.81 | 79.32 |
| 9 | 301.202 | 296.487 | 296.603 | 296.802 | 19.71 | 99.04 |
| 10 | 299.984 | 294.756 | 294.883 | 295.103 | 21.64 | 120.67 |
| 11 | 298.529 | 292.719 | 292.861 | 293.106 | 23.90 | 144.57 |
| 12 | 296.862 | 290.412 | 290.572 | 290.845 | 26.46 | 171.03 |

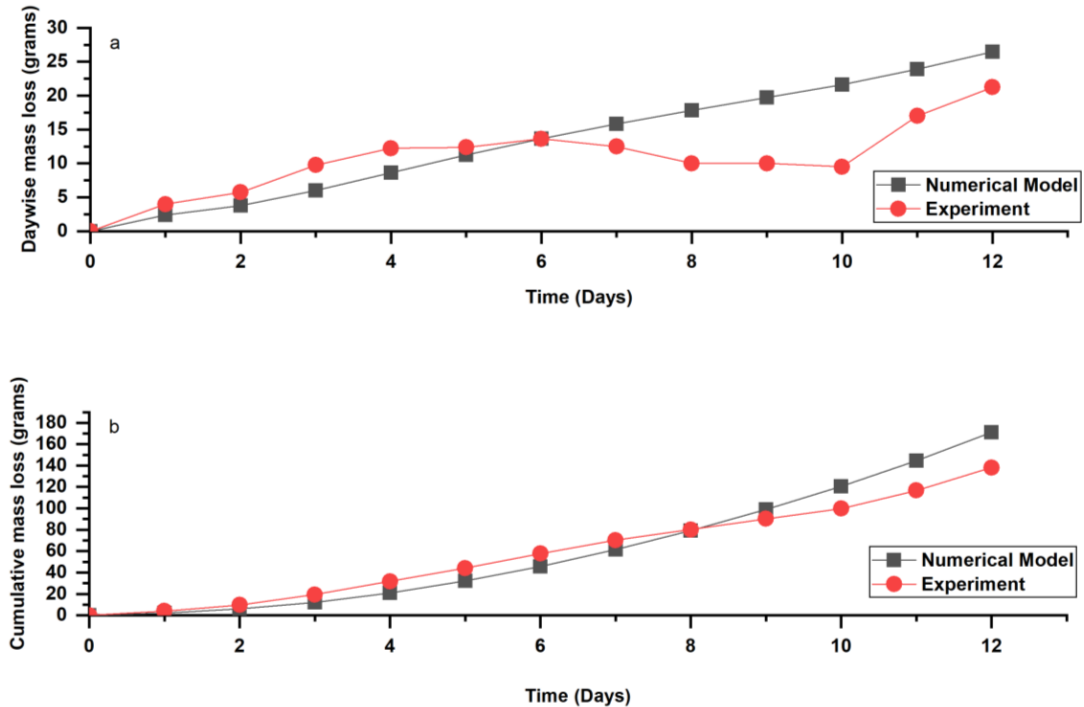

**Fig. 9** Mass loss from numerical model compared with experimental results (a) Daywise, and (b) Cumulative

The daily and cumulative mass loss evaluated from the degraded profile generated using numerical model is in good agreement with the corresponding values evaluated from experimentation [44] as shown in Fig.9. The predicted mass loss by numerical model is on the lower side from mass loss evaluated by varying current in the experiment, and similar observations were made by [54,55]., This validates the numerical model for simulating corrosion in an accelerated corrosion setup with impressed current method. However, relatively larger deviation of mass loss evaluated by the numerical model is observed on $11^{th}$ and $12^{th}$ day, which is because of visible cracks on $11^{th}$ day of accelerated corrosion [44], and was evident by the spike in corrosion current shown in Fig.8. As the PS used in numerical model is evaluated from experimentation on cubes, and cannot reflect the PS levels at the onset of crack, therefore, the model deviates from the experimental result. Further investigation is needed to incorporate the effect of crack on PS variation. Thus, it can be established that the proposed numerical model not only predicts the mass loss with desirable accuracy but also predicts the degradation profile of corroded rebar as a nonlinear time dependent parameter, because of the fact that the mass loss evaluated using numerical model is based on corroded profile of degraded rebar geometry.

## 7 Conclusions

The present work pertains to modelling non uniform time dependent corrosion in 3D domain, analyzing the degraded geometry of corroded rebar, evaluating mass loss and extending the proposed multi-physics numerical model as a tool for real-time simulation of spatial and temporal non-uniformity in corrosion induced deterioration.

The following conclusions can be drawn from the results presented in sections discussed earlier

1. The multi-physics model can be used reliably as a substitute for the accelerated corrosion experiment. The validation shows that the model can predict the non-uniform time dependent mass loss with acceptable accuracy.
2. One of the possible reasons for deviation of weight loss predicted by Faraday's law from experimental method was found to be non-inclusion of differential aeration, resulting in cathodic locations on rebar. Inclusion of these parameters in the proposed numerical method resulted in a much better prediction of weight loss and its localisation.
3. Time dependant weight loss, local current density (day wise), non-uniform degradation of rebar geometry can be obtained at every point on corroding surface using the numerical model. As the mass evaluated by experiment validates the mass evaluated from degraded rebar profile, the degradation profile generated is also validated.
4. Electrolyte conductivity and oxygen diffusivity are taken as function of PS, which are influenced by temperature, humidity, water cement ratio, varying ionic concentration of different chemical species etc. Instead a single parameter such as PS is used to capture the effect, which intern is a function of time and therefore the model can easily be extended for naturally corroding RC members.
5. As PS is continuously monitored and captures the local environment on a real time basis, the model is free from possible deviations arising in prediction models based on climate data such as relative humidity. Also these predictive models have larger possible deviations for long range predictions and are insensitive to local exposure and usage.
6. The model when extended for natural corrosion can also be used for Bayesian updating of predictive models based on climate change. As the model can be updated at each scheduled monitoring of PS and electrolyte conductivity, the models based on relative humidity can be updated with the current level of corrosion evaluated by the model presented in this study.

**Acknowledgements**

The authors acknowledge the support of IMPRINT-2A by Science and Engineering Research Board, Department of Science & Technology, Government of India [grant IMP/2018/001442] for the development of experimental facilities. Authors would like to thank Prof. R K Singh, Dr. Rajeshwara Chary Sriramadasu at IIT Bombay for their suggestions during the course of this research work.

**Disclosure statement**

The authors have no competing interests to declare that are relevant to the content of this article.